\documentstyle[prl,floats,aps,epsf]{revtex}
\begin{document}
\twocolumn
\title{Maximally entangled mixed states in two qubits}
\author{Satoshi Ishizaka and Tohya Hiroshima}
\address{NEC Fundamental Research Laboratories, \\
34 Miyukigaoka, Tsukuba, Ibaraki, 305-8501, Japan}
\date{\today}
\maketitle
\begin{abstract}
We propose novel mixed states in two qubits,
``maximally entangled mixed states'',
which have a property that the amount of entanglement of these states cannot
be increased further by applying any unitary operations.
The property is proven when the rank of the states is less than 4,
and confirmed numerically in the other general cases.
The corresponding entanglement of formation
specified by its eigenvalues gives an upper bound of that
for density matrices with same eigenvalues.
\end{abstract}
\narrowtext
Entanglement (or inseparability) is one of the most striking features of
quantum mechanics
and an important resource for most applications of quantum information.
In quantum computers,
the quantum information stored in quantum bits (qubits)
is processed by operating quantum gates.
Multi-bit quantum gates, such as the controlled-NOT gate,
are particularly important, since these gates can create entanglement between
qubits.
\par
In recent years, it has attracted much attention to quantify the amount of
entanglement, and a number of measures,
such as the entanglement of formation \cite{Bennett96a}
and negativity \cite{Peres96a,Zyczkowski98a},
have been proposed.
When the system of the qubits is in a pure state,
the amount of entanglement can be changed through the gate operations from
zero of separable states to unity of maximally entangled states.
The most quantum algorithms are designed for such ideal pure states.
When the system is maximally mixed, however, we cannot receive any benefit
of entanglement in the quantum computation, since
the density matrix of the system (unit matrix)
is invariantly separable under any unitary transformations or gate operations.
Recently, a question about NMR quantum computation has been proposed
\cite{Braunstein99a},
since the states in the vicinity of the maximally mixed state
are also always separable as is the case of the present NMR experiments.
\par
In all realistic systems, the mixture of the density matrix describing
the qubits is inevitably increased by the 
coupling between the qubits and its surrounding environment.
Therefore, it is extremely important to understand the nature of entanglement
for general mixed states between two extremes of pure states and
a maximally mixed state.
\par
In this paper, we try to answer a simple question how much
the increase of the mixture limits the amount of entanglement 
to be generated by the gate operation or unitary transformation.
We propose a class of mixed states in bipartite 
$2\times2$ systems (two qubits).
The states in this class, {\it maximally entangled mixed states},
show a property of having a maximum amount of entanglement
in a sense that the entanglement of formation (and even negativity)
of these states cannot be increased further by applying any
(local and/or nonlocal) unitary transformations.
The property is rigorously proven in the case that the rank of the states
is less than 4, and confirmed numerically in the case of rank 4.
The corresponding entanglement of formation
specified by its eigenvalues gives an upper bound of that
for density matrices with same eigenvalues.
\par
The entanglement of formation (EOF) \cite{Bennett96a} for a pure state 
is defined as the von Neumann entropy of the reduced
density matrix.
The EOF of a mixed state
is defined as $E_F(\rho)=\min \sum_i p_i E_F(\psi)$, 
where the minimum is taken over all possible decompositions of $\rho$
into pure states $\rho=\sum_i p_i|\psi_i\rangle\langle\psi_i|$.
The analytical form for EOF in 2$\times$2 systems
is given by \cite{Wootters98a}
\begin{equation}
E_F(\rho)=H(\frac{1+\sqrt{1-C^2}}{2}),
\end{equation}
with $H(x)$ being Shannon's entropy function.
The concurrence $C$ is given by
\begin{equation}
C=\max\{0, \lambda_1-\lambda_2-\lambda_3-\lambda_4\},
\end{equation}
where $\lambda$'s are the square root of eigenvalues of $\rho\tilde\rho$
in decreasing order.
The spin-flipped density matrix $\tilde\rho$ is defined as
\begin{equation}
\tilde \rho = \sigma_y^A \otimes \sigma_y^B
\rho^* \sigma_y^A \otimes \sigma_y^B,
\end{equation}
where $^*$ denotes the complex conjugate in the computational basis.
Since $E_F$ is a monotonic function of $C$,
the maximum of $C$ corresponds to the maximum of $E_F$.
\par
The states we propose are that obtained by applying any {\it local} unitary
transformations to
\begin{eqnarray}
M&=&p_1 |\Psi^-\rangle\langle\Psi^-|+p_2 |00\rangle\langle00| \cr
&+&p_3 |\Psi^+\rangle\langle\Psi^+|+p_4 |11\rangle\langle11|,
\label{eq: MEM}
\end{eqnarray}
where $|\Psi^\pm\rangle\!=\!(|01\rangle\!\pm\!|10\rangle)/\sqrt{2}$
are Bell states,
and $|00\rangle$ and $|11\rangle$ are product states orthogonal to
$|\Psi^\pm\rangle$.
Here, $p_i$'s are eigenvalues of $M$ in decreasing order
($p_1 \ge p_2 \ge p_3 \ge p_4$), and $p_1\!+\!p_2\!+\!p_3\!+\!p_4\!=\!1$.
These include states such as
\begin{eqnarray}
\rho&=&p_1 |\Phi^-\rangle\langle\Phi^-|+p_2 |01\rangle\langle01| \cr
&+&p_3 |\Phi^+\rangle\langle\Phi^+|+p_4 |10\rangle\langle10|,
\label{eq: MEM2}
\end{eqnarray}
where $|\Phi^\pm\rangle\!=\!(|00\rangle\!\pm\!|11\rangle)/\sqrt{2}$ are
also Bell states,
and include such that obtained by exchanging
$|\Psi^-\rangle\! \leftrightarrow \!|\Psi^+\rangle$,
$|00\rangle\! \leftrightarrow \!|11\rangle$ in Eq. (\ref{eq: MEM}),
or
$|\Phi^-\rangle\! \leftrightarrow \!|\Phi^+\rangle$,
$|01\rangle\! \leftrightarrow \!|10\rangle$ in Eq. (\ref{eq: MEM2}).
Since entanglement is preserved by local unitary transformations,
all these states have the same concurrence of
\begin{eqnarray}
C^*&=&\max\{0,C^*(p_i)\} \cr
C^*(p_i)&\equiv&p_1-p_3-2\sqrt{p_2p_4}
\end{eqnarray}
The concurrence $C^*$ is maximum among the density
matrices with the same eigenvalues, at least,
when the density matrices have a rank less than 4 ($p_4\!=\!0$).
The proof is as follows:
\par
(1) Rank 1 case ($p_2\!=\!p_3\!=\!p_4\!=\!0$).
In this case, Eq. (\ref{eq: MEM}) is reduced to 
$M=|\Psi^-\rangle\langle\Psi^-|$ which obviously has the
maximum concurrence of unity.
\par
(2) Rank 2 case ($p_3\!=\!p_4\!=\!0$).
Any density matrices of two qubits (not necessarily rank 2) can be
expressed as \cite{Lewenstein98a}
\begin{equation}
\rho=q|\psi\rangle\langle\psi|+(1-q)\rho_{\rm sep},
\end{equation}
where $|\psi\rangle$ is an entangled state and $\rho_{\rm sep}$ is a separable
density matrix.
Convexity of the concurrence \cite{Uhlmann99a} implies that
\begin{equation}
C(\rho)\le q C(|\psi\rangle\langle\psi|)+(1-q)C(\rho_{\rm sep})
=qC(|\psi\rangle\langle\psi|).
\end{equation}
Since $\rho_{\rm sep}$ is a positive operator, $q$ is equal to or less than
the maximum eigenvalue of $\rho$, and thus,
\begin{equation}
C(\rho)\le p_1.
\label{eq: Rank 2 bound}
\end{equation}
The equality is satisfied when $|\psi\rangle$ is a maximally
entangled {\it pure} state and it is an eigenvector of $\rho$ with
the eigenvalue of $p_1$.
The upper bound in Eq. (\ref{eq: Rank 2 bound}) coincides with $C^*$
for $p_3\!=\!p_4\!=\!0$.
\par
(3) Rank 3 case ($p_4\!=\!0$).
Any rank 3 density matrices can be decomposed into two density matrices as
\begin{equation}
\rho=(1-3p_3)\rho_2+3p_3 \rho_3,
\end{equation}
where eigenvalues of $\rho_2$ are
\begin{equation}
\{\frac{p_1-p_3}{1-3p_3},\frac{p_2-p_3}{1-3p_3},0,0\},
\end{equation}
and eigenvalues of $\rho_3$ are $\{1/3,1/3,1/3,0\}$.
According to Lemma 3 in Ref. \onlinecite{Zyczkowski98a},
\begin{equation}
\hbox{Tr}\rho^2 \le \frac{1}{3} \Rightarrow \rho \hbox{~is separable.}
\label{eq: Purity}
\end{equation}
Since the purity of $\rho_3$ is $1/3$, $\rho_3$ is always separable.
Therefore, convexity of the concurrence implies that
\begin{equation}
C(\rho)\le (1-3p_3) C(\rho_2)\le p_1-p_3.
\label{eq: Rank 3 bound}
\end{equation}
Here, we have used that, as shown above, the maximum concurrence of rank 2
density matrices is its maximum eigenvalue.
The upper bound in Eq. (\ref{eq: Rank 3 bound}) again coincides with $C^*$
for $p_4\!=\!0$.
\par
\begin{figure}
\epsfxsize=8.5cm \epsfbox{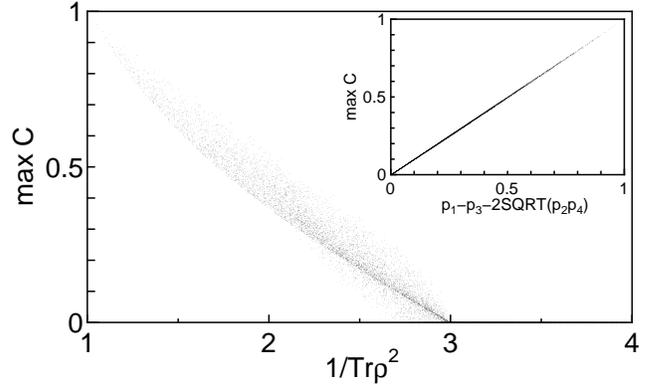}
\caption{
Numerically obtained maximum concurrence for random density matrices as
a function of the participation ratio and $C^*$ (inset).
}
\label{fig: Concurrence}
\end{figure}

In order to check whether $C^*$ is maximum even in general $p_4\!\ne\!0$
cases, we have performed a numerical calculation whose scheme is
similar to that in Ref. \cite{Eisert99a,Zyczkowski98a,Zyczkowski99a}.
We have generated 10,000 density matrices in a diagonal form
with random four eigenvalues \cite{Zyczkowski98a}.
The maximum concurrence has been obtained among 1,000,000 density matrices
generated by multiplying random unitary matrices in the circular unitary
ensemble \onlinecite{Zyczkowski94a} to each of 10,000 diagonal matrices.
The results are shown in Fig.\ \ref{fig: Concurrence} where
the maximum concurrence are plotted as a function of the participation ratio
($R\!=\!1/\hbox{Tr}\rho^2$).
\par
When the density matrix is close to the pure state ($R\!=\!1$),
the maximum concurrence is also close to unity, as expected.
For $R\!\ge\!3$, the states are always separable
(Eq.\ (\ref{eq: Purity}))
and the maximum is zero.
In the region of $1\!<\!R\!<\!3$, the maximum tends to decrease with the
increase of $R$, but the points rather broadly distribute. 
The same data are plotted as a function of $C^*$
in the inset of Fig.\ \ref{fig: Concurrence}.
All points very closely distribute along the straight line of $C\!=\!C^*$,
and none of the points are present on the higher side of the line.
This numerical result strongly support the hypothesis that $C^*$ gives an
upper bound of the concurrence even in the general cases of $p_4\!\ne\!0$.
\par
Accepting the hypothesis implies that all the states satisfying
$C^*(p_i)\!\le\!0$ become automatically separable.
This condition of separability is looser than Eq.\ (\ref{eq: Purity}).
In fact, $C^*(p_i)\!\le\!0$ is only a necessary condition of
$\hbox{Tr}\rho^2\!\le\!1/3$.
Difficulty of the rigorous proof of the hypothesis,
if it is true, might relate to the difficulty of complete understanding of
separable-inseparable boundary in the 15-dimensional space of the density
matrices due to its complex structure.
We emphasize again that the numerical result strongly support
that the hypothesis is true.
\par
It should be noted here that,
when the eigenvalues of a density matrix satisfy a relation,
$C^*$ is indeed maximum even for $p_4\!\ne\!0$.
Any rank 4 density matrices are decomposed as
\begin{equation}
\rho=(p_1-p_3-2\sqrt{p_2p_4})|1\rangle\langle1|
+\rho_4,
\end{equation}
where $|1\rangle$ is an eigenvector of $\rho$,
and eigenvalues of $\rho_4$ (not normalized) are 
$\{p_3+2\sqrt{p_2p_4},p_2,p_3,p_4\}$.
When eigenvalues of $\rho$ satisfies
\begin{equation}
p_3=p_2+p_4-\sqrt{p_2p_4},
\label{eq: Special Condition}
\end{equation}
the purity of (normalized) $\rho_4$ is equal to 1/3 and $\rho_4$ becomes always
separable.
Therefore, using the convexity of the concurrence again, the upper bound of the
concurrence is proven to be $C^*$ for density matrices satisfying
Eq.\ (\ref{eq: Special Condition}).
When $p_2\!=\!p_3\!=\!p_4 (\le\!1/4)$,
$M$ is reduced to the Werner state:
\begin{eqnarray}
M&=&p_1 |\Psi^-\rangle\langle\Psi^-| \cr
&&+\frac{1-p_1}{3}(
|\Psi^+\rangle\langle\Psi^+|
+|\Phi^-\rangle\langle\Phi^-|
+|\Phi^+\rangle\langle\Phi^+|),
\label{eq: Werner}
\end{eqnarray}
whose eigenvalues satisfies Eq.\ (\ref{eq: Special Condition}).
Therefore, EOF of the Werner states cannot be increased further
by any unitary transformations.
\par
\begin{figure}
\epsfxsize=8.5cm \epsfbox{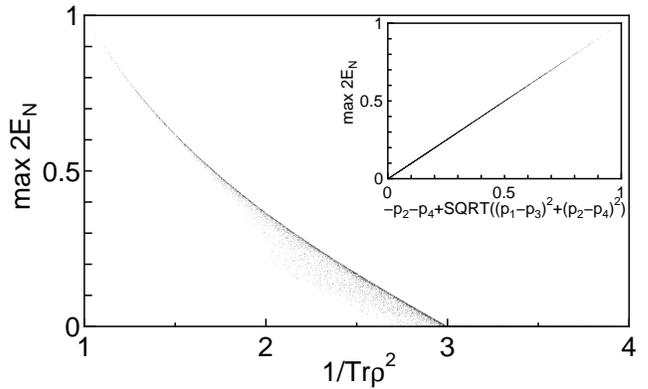}
\caption{
The same as Fig.\ \protect\ref{fig: Concurrence}, but negativity
is plotted.
}
\label{fig: Negativity}
\end{figure}

It is worth to test whether the states we propose have maximum entanglement
in the other entanglement measures.
It has been shown that positive partial transpose is necessary condition for
separability \cite{Peres96a},
and that it is also sufficient condition for $2\times2$
and $2\times3$ systems \cite{Horodecki96a}.
In $2\times2$ systems, when the density matrix is entangled,
its partial transpose has only one 
negative eigenvalue \cite{Sanpera98a}.
The modulus of the negative eigenvalue ($E_N$) is one of entanglement
measures, and twice of $E_N$ agrees with the negativity introduced in
Ref. \onlinecite{Zyczkowski98a}.
We have performed a numerical calculation similar to that for 
Fig.\ \ref{fig: Concurrence}, and obtained maximum of $E_N$ for
10,000 random density matrices.
In the inset of Fig.\ \ref{fig: Negativity}, those
are plotted as a function of
\begin{eqnarray}
2E_N^*&=&\max\{0,2E_N^*(p_i)\}\cr
2E_N^*(p_i)&\equiv&-p_2-p_4+\sqrt{(p_1-p_3)^2+(p_2-p_4)^2},
\end{eqnarray}
which is the negativity of $M$.
None of the points are present on the higher side of a straight line
of $E_N\!=\!E_N^*$ as in the case of the concurrence.
While it has been shown that two measures (EOF and negativity)
do not induce the same ordering of density matrices
with respect to the amount of entanglement \cite{Eisert99a},
the above numerical results suggest that $M$ has the maximum amount of
entanglement in both two measures.
\par
It should be noted here that $E_N^*\!=\!0$ is equivalent to 
$C^*\!=\!0$,
and therefore, the condition of $C^*(p_i)\!\le\!0$,
for which all the states will be separable as mentioned before,
does not contradict to the condition of $E_N^*(p_i)\!\le\!0$.
Further, it will be natural to attribute the upper bound of 
entanglement,
which is well described by $C^*(p_i)$ and $E_N^*(p_i)$,
to the increase of the degree of the mixture of the states.
In this sense, each of $C^*(p_i)$ and $2E_N^*(p_i)$
(both distribute in the range of $[-1/2,1]$)
can be considered as one of measures characterizing the degree of mixture
such as the purity (or participation ratio), von Neumann entropy, 
Renyi entropy, and so on.
\par
Finally, as a simple application of the upper bound of EOF:
\begin{equation}
E_F(\rho) \le H(\frac{1+\sqrt{1-C^*}}{2}),
\label{eq: Upper Bound}
\end{equation}
we consider the situation generating entangled states by using the quantum
gate consisting of two qubits, more concretely a controlled-NOT (CNOT) gate.
In realistic situations, the coupling between
the gate and its surrounding environment is inevitably present.
The entire system consisting of the gate plus its environment happen to
be entangled by the coupling, and the mixture of the reduced density matrix
describing the gate will be inevitably increased.
\par
In order to treat such decohered CNOT gates, we adopt the spin-boson model
\cite{Leggett87a,UWeiss93a}, where each qubit is described as a spin 
$\frac{1}{2}$ system, and the environment is expressed as an
ensemble of independent bosons.
As the model of the CNOT gate, we choose the simplest Hamiltonian:
\begin{equation}
H_G=-\frac{R}{4}(1-\sigma_{cz})\otimes\sigma_{tx},
\end{equation}
where $c$ and $t$ denotes the control-bit and target-bit,
respectively.
The state change after $t\!=\!h/(2R)$ corresponds to the change in the CNOT
operation.
In this paper, we demonstrate two types of gate-environment couplings.
These are
\begin{eqnarray}
H_{GE}^1&=&-\sigma_{cz}\sum_k B_k (a^\dagger_k+a_k), \cr
H_{GE}^2&=&-\frac{1}{2}(1-\sigma_{cz})\otimes\sigma_{tx}\sum_k B_k
(a^\dagger_k+a_k),
\end{eqnarray}
where $a_k$ is an annihilation operator of a boson in the environment.
$H_{GE}^1$ describes the situation that only the control-bit couples with
the environment.
$H_{GE}^2$ may describe the situation that the gate operation is achieved
by irradiating a optical pulse which contains a noise
coherent over the qubits as well as the pulse itself.
For these phase-damping couplings, the time evolution of the reduced density
matrix describing the gate is analytically solved 
\cite{Massimo96a}
by assuming the product initial state for the entire density matrix:
\begin{equation}
\rho_{\rm tot}(0)=\rho(0)\otimes \rho_E,
\end{equation}
where $\rho_E$ is the thermal equilibrium density matrix of the environment.
Since we pay attention only to the generation of the entangled state,
the initial state of the CNOT gate is chosen to be the pure state of
$(|0\rangle_c\!+\!|1\rangle_c)\otimes|0\rangle_t$.
\par
\begin{figure}
\epsfxsize=8.5cm \epsfbox{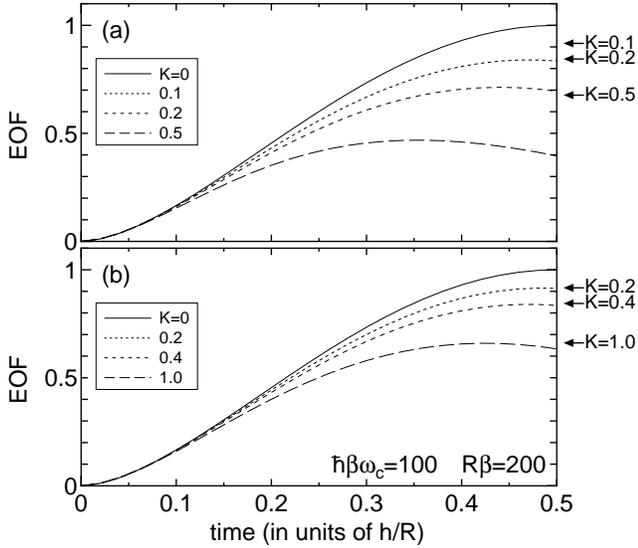}
\caption{
EOF as a function of time in a decohered CNOT gate for several values of
the coupling strength.
$\omega_c$ is the cut-off frequency of the Ohmic environmental mode,
and $\beta$ is the inverse temperature.
(a) $H_{GE}^1$ and (b) $H_{GE}^2$ coupling.
}
\label{fig: Decoherence}
\end{figure}

The time development of EOF for several values of the coupling strength
$K$ ($\propto \sum_k B_k^2 \delta(\omega-\omega_k)$)
are shown in Fig.\ \ref{fig: Decoherence}
for the (a) $H_{GE}^1$ and (b) $H_{GE}^2$ coupling.
The values of the coupling strength $K$ are chosen such that
the values of the fidelity of the output state to the desired state
in the absence of the decoherence
are roughly 0.95, 0.9 and 0.8, which are
common to Figs.\ \ref{fig: Decoherence} (a) and (b).
It is interesting to note that, while the fidelity is the same in
two cases of coupling, the amount of the entanglement is significantly
different.
\par
The upper bound of EOF (Eq.\ (\ref{eq: Upper Bound})) for each coupling
strength is shown by an arrow on the right side of each panel for comparison.
Since EOF of a state has the physical meaning of the asymptotic number
of Bell pairs required to prepare the state by using only local quantum
operations and classical communication, comparing the difference in EOF
will make sense.
In Fig.\ \ref{fig: Decoherence} (a), the EOF is considerably lower than the
corresponding upper bound, while EOF almost agrees with the upper bound in
Fig.\ \ref{fig: Decoherence} (b).
Therefore, with respect to the function generating entangled states,
the performance of the CNOT gate shown in Fig.\ \ref{fig: Decoherence} (b),
is already optimal (or saturated) in a sense that
there is no other way to increase further the amount of the entanglement 
than avoiding the increase of the mixture of the output density matrix,
and thus, avoiding the decoherence itself.
On the other hand, for the CNOT gate shown in Fig.\ \ref{fig: Decoherence} (a),
there is room for improvement of the performance in principle,
although we cannot show the detailed methods here.
\par
To conclude, we propose the mixed states in two qubits,
which have a property that the amount of entanglement of these states cannot
be increased further by any unitary operations.
The property is proven when the rank of the states is less than 4, and
when the states satisfy a special relation such as the Werner states. 
The results of the numerical calculations strongly support a hypothesis
that these mixed states are indeed maximally entangled even in general cases.
It will be extremely important to verify the above hypothesis and
to seek out the {\it maximally entangled mixed states} as well as
the measure in larger dimensional systems,
for understanding the nature of entanglement of general mixed states
and for the progress of the quantum information science and its applications.

\begin{thebibliography}{10}

\bibitem{Bennett96a}
C.~H. Bennett, D.~P. DiVincenzo, J.~A. Smolin, and W.~K. Wootters, Phys. Rev. A
  {\bf 54},  3824  (1996).

\bibitem{Peres96a}
A. Peres, Phys. Rev. Lett. {\bf 76},  1413  (1996).

\bibitem{Zyczkowski98a}
K. \.Zyczkowski and P. Horodecki, Phys. Rev. A {\bf 58},  883  (1998).

\bibitem{Braunstein99a}
S.~L. Braunstein {\it et~al.}, Phys. Rev. Lett. {\bf 83},  1054  (1999).

\bibitem{Wootters98a}
W.~K. Wootters, Phys. Rev. Lett. {\bf 80},  2245  (1998).

\bibitem{Lewenstein98a}
M. Lewenstein and A. Sanpera, Phys. Rev. Lett. {\bf 58},  2261  (1998).

\bibitem{Uhlmann99a}
A. Uhlmann, quant-ph/9909060  .

\bibitem{Eisert99a}
J. Eisert, J. Mod. Opt. {\bf 46},  145  (1999).

\bibitem{Zyczkowski99a}
K. \.Zyczkowski, Phys. Rev. A {\bf 60},  3496  (1999).

\bibitem{Zyczkowski94a}
K. \.Zyczkowski and M. Ku\'s, J. Phys. A {\bf 27},  4235  (1994).

\bibitem{Horodecki96a}
M. Horodecki, P. Horodecki, and R. Horodecki, Phys. Lett. A {\bf 223},  1
  (1996).

\bibitem{Sanpera98a}
A. Sanpera, R. Tarrach, and G. Vidal, Phys. Rev. A {\bf 58},  826  (1998).

\bibitem{Leggett87a}
A.~J. Leggett {\it et~al.}, Rev. Mod. Phys. {\bf 59},  1  (1987).

\bibitem{UWeiss93a}
{\em Quantum Dissipative Systems}, edited by U. Weiss (World Scientific,
  Singapore, 1993).

\bibitem{Massimo96a}
G.~M. Palma, K.-A. Suominen, and A.~K. Ekert, Proc. R. Soc. Lond. A {\bf 452},
  567  (1996).

\end{thebibliography}

\end{document}